\documentclass[%
 reprint,
 amsmath,amssymb,
 aps,soul,
pre,
]{revtex4-2}

\usepackage{graphicx}
\usepackage{dcolumn}
\usepackage{bm}
\usepackage{xcolor}
\usepackage{url}

\usepackage{lipsum}
\makeatletter
\newcommand*{\balancecolsandclearpage}{%
  \close@column@grid
  \cleardoublepage
  \twocolumngrid
}

\renewcommand{\selectlanguage}[1]{}

\begin{document}

\preprint{APS/123-QED}

\title{Kinetic renormalization of auroral turbulence}


\author{Magnus F Ivarsen}
\email{Contact: magnus.fagernes@gmail.com}
\altaffiliation[Also at ]{The European Space Agency Centre for Earth Observation, Frascati, Italy}
\affiliation{Department of Physics and Engineering Physics, University of Saskatchewan, Saskatoon, Canada}

\author{Kaili Song}
\affiliation{Physics Department, University of New Brunswick, Fredericton, Canada}

\author{Luca Spogli}
\affiliation{Istituto Nazionale di Geofisica e Vulcanologia, Rome, Italy}

\author{Jean-Pierre St-Maurice}
\altaffiliation[Also at ]{Department of Physics and Astronomy, University of Western Ontario, London, Canada}
\author{Brian Pitzel}
\author{Saif Marei}
\affiliation{Department of Physics and Engineering Physics, University of Saskatchewan, Saskatoon, Canada}

\author{Devin R Huyghebaert}
\altaffiliation[Also at ]{Department of Physics and Engineering Physics, University of Saskatchewan, Saskatoon, Canada}
\affiliation{Leibniz Institute of Atmospheric Physics, K\"{u}hlungsborn, Germany}

\author{Yangyang Shen}
\affiliation{Department of Earth and Space Sciences, University of California, Los Angeles, USA}

\author{Satoshi Kasahara}
\author{Kunihiro Keika}
\affiliation{Department of Earth and Planetary Science, University of Tokyo, Tokyo, Japan}

\author{Yoshizumi Miyoshi}
\author{Tomo Hori}
\author{Atsuki Shinbori}
\author{Kazuhiro Yamamoto}
\affiliation{Institute for Space-Earth Environmental Research, Nagoya University, Nagoya, Japan}

\author{David R Themens}
\affiliation{School of Engineering, University of Birmingham, Birmingham, UK}

\author{Yoichi Kazama}
\author{Shiang-Yu Wang}
\affiliation{Academia Sinica Institute of Astronomy and Astrophysics, Taipei, Taiwan}

\author{Ayako Matsuoka}
\affiliation{Data Analysis Center for Geomagnetism and Space Magnetism, Kyoto University, Kyoto, Japan}

\author{Iku Shinohara}
\author{Takefumi Mitani}
\author{Takeshi Takashima}
\affiliation{Institute of Space and Astronautical Science, Japan Aerospace Exploration Agency, Sagamihara, Japan}

\author{Shoichiro Yokota}
\affiliation{Department of Earth and Space Science, Osaka University, Toyonaka, Japan}

\author{P. T. Jayachandran}
\affiliation{Physics Department, University of New Brunswick, Fredericton, Canada}

\author{Glenn C Hussey}
\affiliation{Department of Physics and Engineering Physics, University of Saskatchewan, Saskatoon, Canada}


\begin{abstract}
Driven-dissipative systems often exhibit self-organization in the form of coherent dissipative structures. However, observing such critical states in natural plasmas remains elusive, leading to the traditional view that the fine structure of Earth's auroral ionosphere is shaped by local turbulent flows. Here we report the discovery of a self-organizing regime in Earth's ionosphere. We identify this by modeling the sum of saturation electric fields in the turbulent auroral electrojets as a stochastic variable that renormalizes into noise-enabled transport, via explicitly derived Bohm diffusion. This constitutes an effective field-theory for Farley-Buneman turbulence in the Martin-Siggia-Rose formalism for renormalization group theory, for which we provide strong empirical evidence. Using a composite radar-GPS power spectrum of plasma turbulence, we resolve a scale-invariant cascade that exhibits a characteristic kinetic Alfv\'{e}n $k^{-8/3}$-signature across four orders of magnitude in $k$. What is more, a large statistical analysis of how the turbulence responds to magnetospheric driving reveals a clear tendency for the observed number density of turbulent waves to scale linearly with driving power, matching the predictions made by our field theory's overdamped equations of motion, which offer closed-form calculations of macroscopic transport relations that are uniquely suitable for sub-grid parameterization in space weather modeling. This establishes geospace storms as opportunities to observe non-equilibrium phase transitions imposing global constraints on collision-dominated systems.
\end{abstract}

\maketitle


\section{\label{sec:intro}Introduction}

In Earth's lower ionosphere, collisions between charged particles and neutral molecules become so frequent that  the ratio of collision to cyclotron frequency is large enough to demagnetize ions below 120 km and electrons below 80 km. The resulting difference in charge carrier mobility between 80 and 120 km allows strong electric fields in the aurora to drive  Hall currents -- a reservoir of free energy for electrostatic instabilities \cite{fejerIonosphericIrregularities1980,hubaIonosphericTurbulenceInterchange1985}.  For plasma drifts below around 400~m/s, the currents remain laminar, owing to ordinary diffusion. 
However, when the electron drift exceeds the  ion acoustic speed, $C_s$, the laminar current becomes unstable to electrostatic density fluctuations,  triggering the Farley-Buneman (FB) instability through the ion inertia excited by the motion of the irregularities, which is driven by the electron plasma ($E\times B$) drift. \cite{farleyPlasmaInstabilityResulting1963,bunemanExcitationFieldAligned1963} Given fast drifts, the resulting state of the plasma is one of intense, broad-spectrum electrostatic turbulence \cite{sato_stabilization_1972,sudan_theory_1977,keskinen_nonlinear_1981} that strongly scatters radio waves (`radio aurora') \cite{hultqvistRadioAurora1964,hysellRadarAurora2015} and generates an increase in the overall Joule heating rate through the  nonlinear dissipation of electrostatic structures. \cite{st-mauriceRevisitingBehaviorERegion2021} The dissipation occurs soon after the unstable structures have reached maximum amplitude, having taken as much free energy as possible from the system, thereby bringing the electric field to a point of marginal stability. \cite{st.-mauriceNewNonlinearApproach2001}  The dissipation comes from the inevitable creation of increasingly important wave electric field components along the magnetic field, which erode the structures \cite{drexlerNewInsightsNonlocal2002,oppenheim_kinetic_2013} and thereby returns the energy to the particles via acceleration. For very strong ambient Hall currents, the additional heating rate from the energy going through FB waves becomes highly relevant through large increases in the ambient electron temperature above 105 km \cite{schlegelAnomalousHeatingPolar1981, st-maurice_nonlocal_1985, st-mauriceElectronHeatingPlasma1990,hamza_mode-coupling_1998,st.-mauriceNewNonlinearApproach2001,st-mauriceRevisitingBehaviorERegion2021}.

The foregoing describes a dynamic and complex conversion of  ambient Hall currents into electrostatic energy, as demonstrated through in-situ measurements \cite{pfaff_electric_1987, kelley_electric_1986, kelley_condor_1986}, ground-based radar studies, \cite{st.-mauriceFirstResultsObservation1989, fosterAspectAngleVariations1992, fosterPhaseVelocityStudies1992} and hybrid fluid- and particle-simulations \cite{oppenheim_hybrid_1995, oppenheim_saturation_1996, oppenheim_kinetic_2013}. Recently, radar studies have also { shown that the  turbulence  is triggered preferentially } on the edges of the precipitation regions, \cite{bahcivanObservationsColocatedOptical2006, huyghebaertPropertiesICEBEARERegion2021} where the electric field intensifies \cite{fujii_reformulation_2011,ivarsen_turbulence_2024}.

Building on this description of externally driven FB turbulence, we surmise a state of criticality, where the response of the non-laminar electrojets around diffuse aurorae is \textit{saturated}, and the system is forced into an overdamped state. To capture the saturation dynamics, we adopt a formalism consistent with Refs.~[\cite{hamza_turbulent_1993}] and [\cite{st.-mauriceTheoreticalFrameworkChanging2016}]: {\textit{(1)} the relevant driving term is controlled by a  Doppler-shifted eigenfrequency $\omega' = \omega - \mathbf{k} \cdot \mathbf{V}_d$ that plays an important role after we consider that \textit{(2)} the instability is regulated by an effective damping term, $\gamma_\text{eff}$,  which forces the growth rate  to go from large at early stages where the structures are unstable. After that, the growth rate of single structures goes to zero owing to the nonlinear association of the wave field with the perturbed density, in reaction at first. \cite{st.-mauriceNewNonlinearApproach2001} In their final stage of evolution, the growth of the wave-field, along the magnetic field, damps the unstable structures and dissipates them.}

In specific terms, the dynamics of the density fluctuations in the frame of the drifting plasma are governed by a dispersion relation, that favors growth at first and then becomes highly dissipative, all through the changes in the nonlinear evolution of the eigenfrequency, as given by, e.g., Ref.~[\cite{hamza_turbulent_1993}],
\begin{equation} \label{eq:omega}
{\omega}^{2} - k_{\perp}^{2}C_{s}^{2} + i\omega' \gamma_\text{eff} = 0,
\end{equation}
where $k_\perp$ is the perpendicular wavenumber and $\gamma_\text{eff}$ is large, owing to the small magnitude of Pedersen currents relative to electron Hall currents.  

By applying renormalization group (RG) theory, we demonstrate that the turbulent ionosphere flows to a macroscopic effective field theory described by a deterministic advection-diffusion equation. In this formulation, the spatial dynamics are governed by an effective diffusion tensor,
\begin{equation} \label{eq:deff}
\mathbf{D}_\text{eff} = \gamma_\text{eff}^{-1}(C_s^2\, \mathbb{I} - \mathbf{V}_d \otimes \mathbf{V}_d), \end{equation} where $C_s$ is the acoustic speed and $\mathbf{V}_d$ is the saturating drift velocity. Here, the instability manifests as negative diffusion \cite{ashinsky_nonlinear_1988} instead of wave-growth, triggered when drifts surpass the ion acoustic speed ($D_\parallel \sim C_s^2 - V_d^2 < 0$). This unstable growth occurs only along the mean drift, while transverse modes, the overdamped secondary waves, \cite{hamza_turbulent_1993} remain subject to positive effective diffusion $D_{\perp} \sim C_s^2$. The instability grows until it reduces the driving field $\mathbf{V}_d$, flipping the diffusion from negative to positive and ensuring marginal stability.


Solving the resulting advective-diffusive equation of motion for the renormalized FB waves yields an Adler-Ohmic bifurcation in the system, where magnetospheric forcing elicits a linear (Ohmic) response in the degree of proliferation of FB turbulence. The rate of auroral very high-frequency (VHF) radar echo detection scales linearly with driver power $n \propto (2\lambda_0^2)^{-1}\Delta E$, where $\lambda_0$ is the  renormalized, anomalous diffusion coefficient (see Section~\ref{sec:tensorhome} in the Methodology). 
We demonstrate the validity of the theory by showing that, during intense diffuse aurorae, the ionosphere-magnetosphere coupling exhibits the theorized Adler-Ohmic bifurcation, \cite{adler_study_1946} with an Ohmic response where E-region turbulence intensity $n$ scales linearly with electromagnetic power input: $n \propto \Delta E$, but will only do so when $\Delta E$ exceeds a characteristic threshold wave power $\Delta E_c$, measured by the Arase spacecraft,\cite{miyoshi_geospace_2018} We explicitly predict that the linear relationship follows from the inability of the penetrating electric fields to accelerate the waves (coming up against the acoustic speed), forcing the number of turbulent parcels (proliferation) to track the driver. The ionosphere around diffuse aurorae thus self-organizes into a driven-dissipative boundary, preserving the topology of the magnetospheric turbulence while optimizing energy dissipation.

\begin{figure}
    \centering
    \includegraphics[width=.495\textwidth]{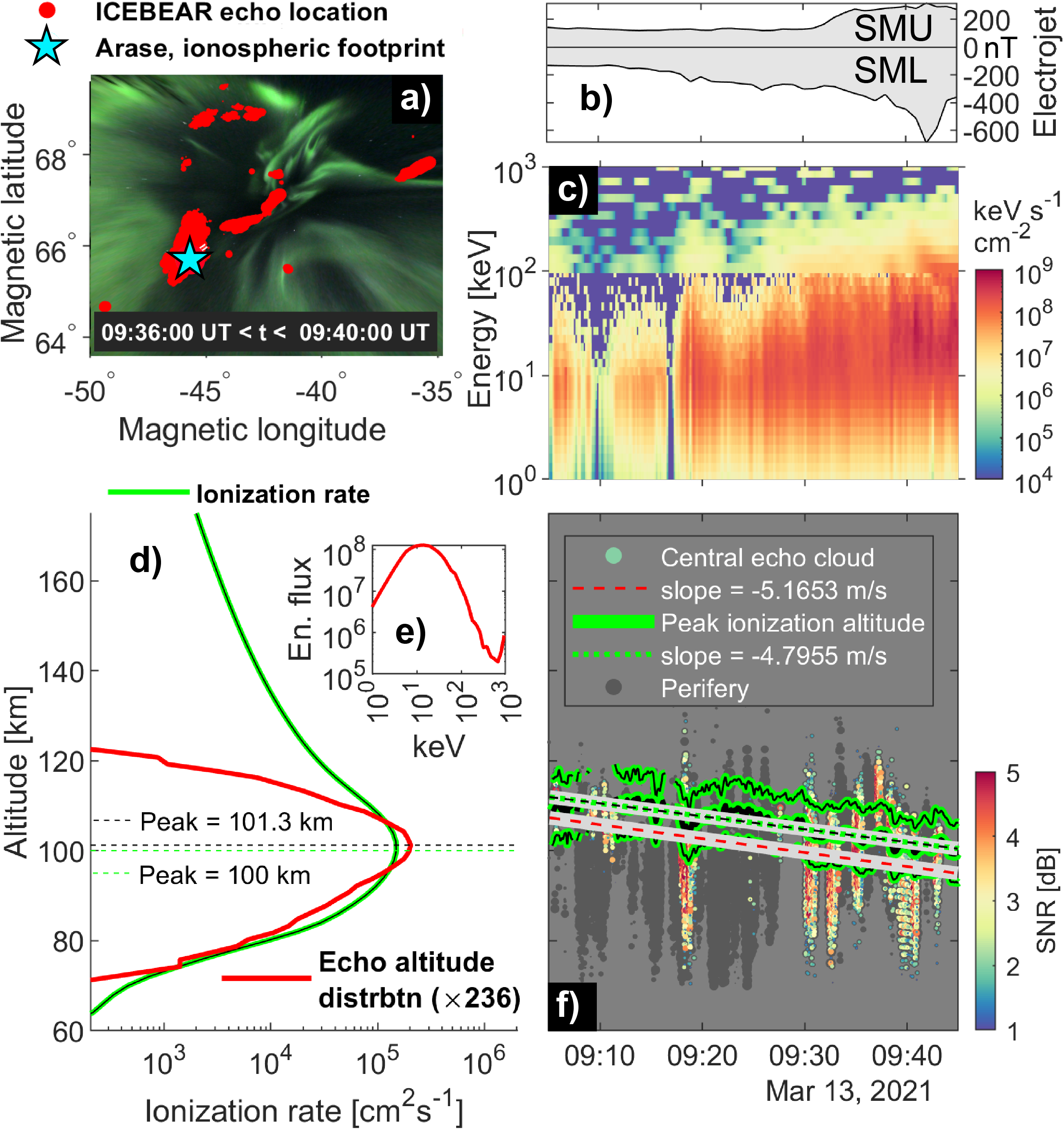}    
    \caption{\textbf{A space-ground conjunction  between the \textsc{icebear} radar and the inner-magnetosphere spacecraft Arase,} \cite{miyoshi_erg_2018,miyoshi_geospace_2018} that took place on 13 March 2021. \textbf{Panel a)} shows auroral images from the \textsc{tre}x \textsc{rgb} \cite{gilliesApparentMotionSTEVE2020} all-sky-imager at Rabbit Lake, with radar echo locations in red and Arase's ionospheric footprint as a cyan star. \textbf{Panel b)} shows the upper and lower envelope of the auroral electrojet index, that is, the SuperMAG electrojet (SME) index, \cite{newellEvaluationSuperMAGAuroral2011} demonstrating the onset of a moderate magnetospheric substorm. \cite{newellSubstormMagnetosphereCharacteristic2011} \textbf{Panel c)} shows the combined precipitating electron energy flux through the LEP-e, \cite{kazama_low-energy_2017-1} MEP-e, \cite{kasahara_medium-energy_2018} and HEP \cite{mitani_high-energy_2018} instruments (pitch-angles lower than 5$^\circ$ and $10^\circ$), which we take as a proxy for the real precipitating energy flux). \textbf{Panel d)} compares the inferred ionization altitude profile from Arase (green), based on the accumulative precipitating energy flux through the interval, with the altitude distribution of radar echoes superposed (red line), ignoring echoes with extreme azimuths, whose altitudes are anomalous. \cite{ivarsenAlgorithmSeparateIonospheric2023} \textbf{Panel e)} shows the median precipitating electron spectra, while \textbf{panel f)} show altitude-time-intensity point-clouds of echoes, color-coded by signal-to-noise ratio (SNR). Green and black lines indicate the peak ionization altitude (thick line) and a single standard deviation (thin lines). Echoes colored dark grey are observed outside of the radar's central field-of-view, whose altitudes are anomalous. \cite{ivarsenAlgorithmSeparateIonospheric2023}
    }
    \label{fig:ex}
\end{figure}

\section{Results}

The kinetic energy of the particle precipitation in diffuse and pulsating auroral patches, created when wave-particle interactions leads to pitch-angle scattering of hot electrons in Earth's radiation (Van Allen) belts, \cite{thorne_scattering_2010,kasaharaPulsatingAuroraElectron2018} can be exceedingly high, with significant precipitation at energies exceeding $30$~keV. \cite{nishimuraDiffusePulsatingAurora2020} This ionizes the ionosphere at low altitudes,  \cite{fangParameterizationMonoenergeticElectron2010,miyoshi_energetic_2015,miyoshi_penetration_2021} where the FB instability becomes modulated by ohmic wave-heating  \cite{dimant_physical_1997,st.-mauriceTheoreticalFrameworkChanging2016} and the electron thermal instability,  \cite{dimantIonThermalEffects2004} meaning electrojet turbulence will effectively accompany particle precipitation throughout the E-region. Figure~\ref{fig:ex} shows measurements that confirm the foregoing, performed during the onset of a magnetospheric substorm, which was captured by a conjunction between the 3D VHF radar \textsc{icebear} and the inner-magnetosphere spacecraft Arase. The altitude-distribution of backscatter turbulence echoes matches that of the ionization altitudes with high accuracy (Figure~\ref{fig:ex}d, f), with a strikingly similar downward speed in the peak altitudes ($\sim5$~m/s).

\begin{figure*}
    \centering
    \includegraphics[width=\textwidth]{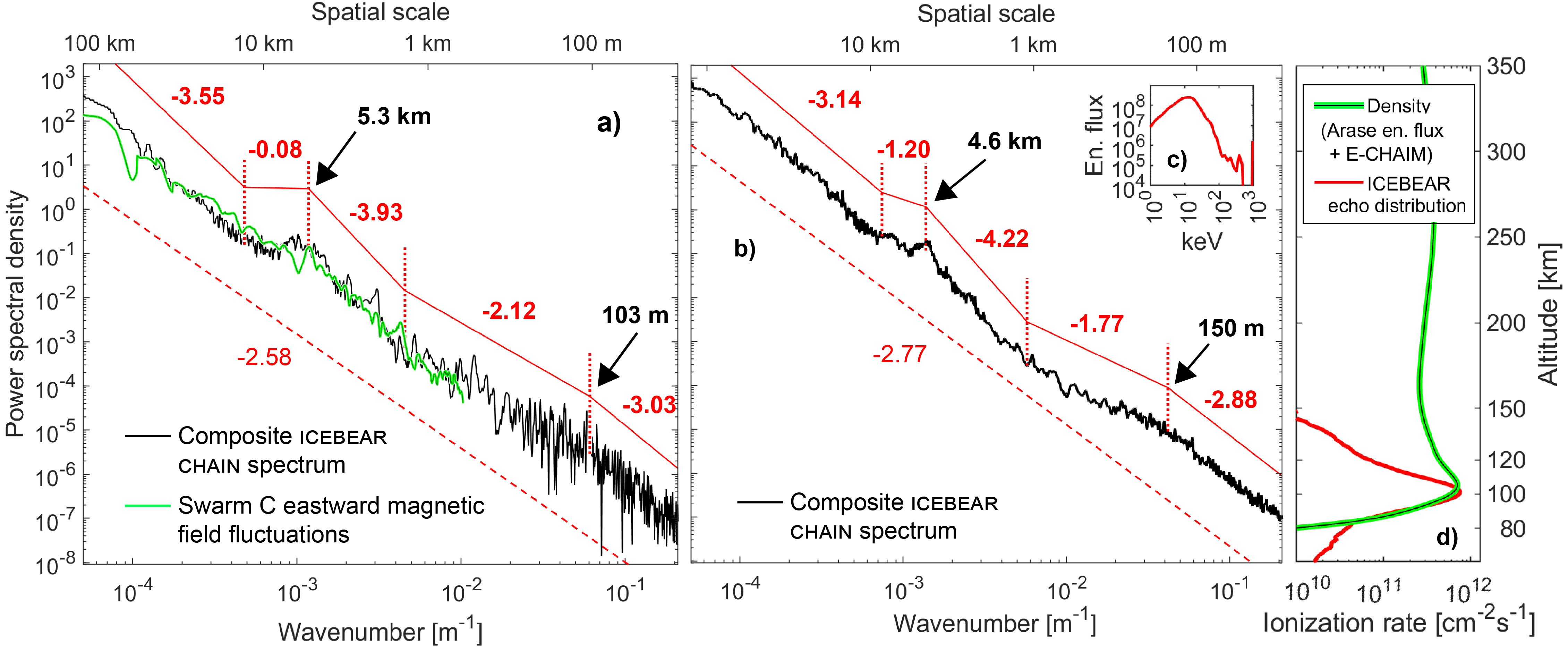}
    \caption{\textbf{Space-ground-ground conjunctions between \textsc{icebear}, \textsc{chain}, Swarm, and Arase. Panel a)} shows the average composite spectrum for the five-minute interval starting at 11:06~UT on 6 May 2023 (Black line). The normalized spectrum of eastward magnetic fluctuations measured by Swarm~C is shown with a green line. A five-component piecewise log-log linear fit is shown above the spectra (solid red line) while a single-slope fit is shown below the spectrum (dashed red line). Spectral indices are indicated in red lettering while two prominent spatial scales (break-points) are indicated with black lettering. \textbf{Panel b)} shows the composite \textsc{icebear}-\textsc{chain} spectrum measured during the 17-minute interval starting at 13:02~UT on 2 August 2023, akin to panel a). \textbf{Panel c)} shows the average precipitating electron spectrum measured by Arase during the interval, while \textbf{panel d)} shows (in green) the plasma density column created by the particles, using the Empirical Canadian High Arctic Ionospheric Model (E-CHAIM) \cite{themensEmpiricalCanadianHigh2017} to incorporate the effects of solar extreme-ultraviolet photoionization. Whereas the overall shape is steep, there are recurring spectral \textit{break-points} at kilometer and decameter scales in \textbf{panels a, b)}. These exhibit distinct inertial and dissipative components, clear indications of a classical instability mechanism (e.g., Ref.~[\cite{mounirSmallscaleTurbulentStructure1991}]), consistent with gradient-drift structures, \cite{greenwaldDiffuseRadarAurora1974} and the low-altitude echo distribution in \textbf{panel d)} f is consistent with Ohmic wave-heating \cite{dimant_physical_1997} or thermal-electron instabilities, \cite{dimantIonThermalEffects2004} as well as conventional electrojet turbulence. 
    See the Supplementary Materials for further details of the conjunctions.
    }
    \label{fig:swarmcomp}
\end{figure*}

Figure~\ref{fig:swarmcomp}a) shows the triple conjunction between \textsc{icebear}, \textsc{chain}, and the Swarm~A \& C satellites. \cite{friis-christensenSwarmConstellationStudy2006,wood_variability_2022} Panel a) shows the composite spectrum of plasma structuring in the E-region during the conjunction. In green, we show the spectrum of filamentary field-aligned currents, measured by Swarm~C. The exponent of the fitted power laws, referred to as spectral index, tracks energy dissipation, \cite{ivarsenDirectEvidenceDissipation2019} and we observe that the shape-wise agreement between the spectra is excellent, echoing recent conjunctions of a similar nature, \cite{ivarsenDistributionSmallScaleIrregularities2023,ivarsenMeasuringSmallscalePlasma2023,ivarsen_turbulence_2024-1} and consistent with the filamentary nature of the field-aligned currents observed around diffuse aurorae. \cite{gilliesSwarmObservationsFieldaligned2015}

Next, in Figure~\ref{fig:swarmcomp}b--d), we summarize the observations from an extended space-ground-ground conjunction that took place between \textsc{icebear}, \textsc{chain}, and the Japanese inner-magnetosphere spacecraft Arase, the latter of which observed an intense flux of energetic particles precipitating directly towards a region where \textsc{icebear} concurrently recorded some four million individual echo locations. The inferred plasma density altitude profile matches the echo distribution, and the observed GPS- and radar-spectra co-evolve (see Supplementary Materials).  

Figure~\ref{fig:kinetic}a--d) present statistics of the spectra, demonstrating a clear and consistent tendency for a value around $-8/3$ as the most probable measurement value overall, and a relatively narrow spread around that value. The echo altitude distribution from the 10 events in Figure~\ref{fig:kinetic}c) peaks at 102~km. Using this altitude for the GPS pierce points gives us a phase screen drift of 500 m/s, coinciding with the peak of the median Doppler speed distribution for the 10 events (Figure~\ref{fig:kinetic}b). The FB wave phase speed saturates at the ion acoustic speed. \cite{fosterSimultaneousObservationsEregion2000,oppenheim_kinetic_2013,chauUnusualRegionFieldaligned2016} or around 500~m/s, demonstrably in agreement with Figure~\ref{fig:kinetic}b). This kinetic constraint confirms that the GPS scintillations took place in the E-region, a notion supported by recent statistical studies. \cite{blinstrubas_comparative_2025}

\begin{figure*}
    \centering
    \includegraphics[width=\textwidth]{ohmic-02.png}
    \caption{
    \textbf{Panel a)} compares the \textsc{icebear}-\textsc{chain} spectra (green line) with a statistical aggregate of 7,700 \textsc{icebear} echo clustering spectra (red line) observed during 2020, 2021 (the database that was analyzed in Ref.~[\cite{ivarsenMeasuringSmallscalePlasma2023}]). \textbf{Panel b) } shows the median distributions in HR radar Doppler speeds, while \textbf{panel c)} treats echo  altitudes. A green line in panel c) indicates the assumed GPS pierce-point locations (with shaded green region giving upper/lower quartile distributions), whereas a green line in panel b) shows the derived phase screen speed. \textbf{Panel d)} shows all 10 composite \textsc{icebear}-\textsc{chain} spectra (solid black line), with a linear log-log fit (black dashed line) and a -8/3-slope (red dashed line) indicated.   \textbf{Panel e)} aggregates the \textsc{icebear} echo detection rate, for some 1.2~million one-second intervals (containing a total of 272~million echoes), while \textbf{Panel f)} aggregates Arase-observed wave power (integrated wave spectra for frequencies $f>0.1f_e$, $f_e$ being the electron cyclotron frequency) for some 530,000 one-second wave spectra, measured using the Plasma Wave Experiment on board Arase \cite{kasahara_plasma_2018}, using measurements from when the satellites was within $5^\circ$ off magnetospheric equator, and during magnetic local times \cite{bakerNewMagneticCoordinate1989} between 22h and 06h, with data collected on days when the radar was operational (between January 2020 and June 2023). In both panels e) and f), red triangles represent the median value inside 15 logarithmically spaced geomagnetic activity bins (using the SME-index, \cite{newellEvaluationSuperMAGAuroral2011} or auroral electrojet index). Vertical errorbars denote upper/lower quartile distributions. Linear fits are indicated, by non-linear least squares minimization of the root-mean-square error. \textbf{Panel g)} shows the same bins in a scatterplot, with a black dashed line denoting $n\propto\Delta E^\beta$ (see Eq.~\ref{eq:theory}), with $\beta=0$ when $\Delta E<\Delta E_c$ and $\beta=0.98\pm0.09$ otherwise, determined through non-linear least squares minimization, where $\Delta E_c\sim10^{-4}$~mV$^2$m$^{-2}$ represents the critical wave power. Statistical error margin is posted showing 95-percent confidence intervals for $\beta$ (3-sigma), using a Monte Carlo-based bootstrapping routine with 500 iterations and a randomized data selection for each iteration.}
    \label{fig:kinetic}
\end{figure*}

Next, we shall describe the energy transfer of ionosphere-magnetosphere coupling during such observations of intense auroral plasma turbulence, by statistically aggregating \textsc{icebear} and Arase observations. Figure~\ref{fig:kinetic}e--g) we show median values of the VHF radar echo detection rate $n$ and the total integrated wave power $\Delta E$, for 15 geomagnetic activity bins (see Ref.~[\cite{ivarsen_characteristic_2025}]). In Figure~\ref{fig:kinetic}e), we observe a distinct elbow in the curve at an auroral electrojet value of around 150~nT, below which the echo detection rate is flat (meteor trail echoes \cite{oppenheimPlasmaInstabilitiesMeteor2003,husseyComparisonNorthernHemisphere2000,ivarsenAlgorithmSeparateIonospheric2023}) For bins above this elbow, the bin-median echo detection rate is near-perfectly correlated with the auroral electrojets ($\rho=0.99$), and at the same threshold, we observe that the wave power stabilizes against the electrojet index {$\rho=0.95$). We label this elbow with the critical wavepower $\Delta E_c$. When the magnetospheric wave power exceeds this threshold, the two quantities exhibit a similar response to enhancements in the electrojets, facilitating the linear ($\beta=0.98\pm0.09$) relationship shown in Figure~\ref{fig:kinetic}g).

\section{Discussion}

The statistical aggregates in Figure~\ref{fig:kinetic} clearly demonstrate that the composite E-region power spectra tend to be \textit{steep}, peaking at $\alpha\approx-8/3$, and that the VHF backscatter echo rate statistically exhibits an overall Ohmic (linear) response to magnetospheric forcing, the latter approximated by observed wave power near the radiation belts.


The specific physical mechanism enforcing the linear scaling between the driver and the proliferation of FB turbulence ($\beta \approx 1$) can be derived from the non-linear evolution of finite plasma structures. Following Ref.~[\cite{st.-mauriceNewNonlinearApproach2001}], we consider that an initial density perturbation evolves as a finite structure that generates internal polarization electric fields. These fields oppose the driver, effectively reducing the internal electric field until the drifts barely exceed the threshold condition ($C_s$). \cite{oppenheim_kinetic_2013} This feedback mechanism explains the pervasive observation of phase velocities saturated near the ion-acoustic speed. \cite{fosterSimultaneousObservationsEregion2000,oppenheim_kinetic_2013}
Crucially, the aggregate intensity of the turbulence is determined by the density of these structures within the unstable volume. We posit a space-time equivalence where the number of active scatterers, $n$, is proportional to the linear growth rate $\gamma_{lin}$.  \cite{drexlerNewInsightsNonlocal2002} Since the fluid growth rate scales as $\gamma_{lin} \propto (V_{d}^2 - C_s^2)$, and the drift velocity $V_{d}$ is proportional to the external electric field $E$, the strongly driven regime ($V_{d}^2 \gg C_s^2$) yields a scatterer density $n \propto E^2$ (see Section~\ref{sec:rg:kinetic}). The linear response is therefore a direct consequence of the instability actively populating the volume in proportion to the available free energy.

A rigorous application of renormalization group theory \cite{wilson_renormalization_1971} completes the foregoing phenomenological description (see Section~\ref{sec:rg}). By reconstructing an action in the path-integral formulation, working backwards from the dispersion relation Eq.~(\ref{eq:omega}), we apply a spacetime rescaling ($x' \to b^{-1}x$), after which the inertial operator responsible for acceleration becomes irrelevant, decaying as $b^{-2}$ at macroscopic scales. Consequently, the system flows inevitably toward a stable, \textit{overdamped} fixed point (dynamic exponent $z=2$).

The overdamped equations of motion define an effective field theory with an infrared cutoff (correlation scale) given by (see Section~\ref{sec:rg:scale}):
\begin{equation} \label{eq:cross}
L_c = \frac{2\sqrt{2}\pi D_\text{eff}}{V_{d}-C_s},
\end{equation}
where $V_d$ is the saturated drift. $L_c$ features a divergence at the threshold ($C_s$), characteristic of a continuous phase transition, \cite{hohenberg_theory_1977} but the excess drift saturates at around $1.1C_s$ inside the turbulence, \cite{hamza_turbulent_1993,oppenheim_kinetic_2013} keeping Eq.~(\ref{eq:cross}) finite.

The turbulent amplitude field $\psi=\delta N_e/N_{e,0}$ is governed by an effective transport equation. Here, $V_d > C_s$ triggers turbulence growth via negative diffusion. \cite{ashinsky_nonlinear_1988} The net stability condition decomposes into a driving term (source) and a saturation term (sink) (Eq.~\ref{eq:transport1}):
\begin{equation} \label{eq:transport2}
    D_{\parallel} \approx \gamma_\text{eff}^{-1}\left[ (C_s^2 - V_{d0}^2) + 2\tilde{\alpha} V_{d0}^2 |\psi|^2\right].
\end{equation}

Marginal stability implies a balance where the net diffusion vanishes ($D_\parallel \to 0$), meaning the anomalous saturation term must exactly cancel the unstable driver. Physically, this corresponds to an anomalous electric field,
\begin{equation} \label{eq:eanom}
E_\text{anom} \approx (V_{d0}-C_s)B,
\end{equation}
that acts as a macroscopic drag force. This drag underpins the resistive non-linear currents \cite{dimantEffectsElectronPrecipitation2021} and wave-heating \cite{st.-mauriceTheoreticalFrameworkChanging2016} observed \emph{in-situ} and in kinetic simulations.
Figure~\ref{fig:kinetic}g) offers clues as to how this drag manifests in the dynamics. The observed linear relationship $n \propto \Delta E$ implies that the \textit{number density of turbulent waves} obeys Ohm's law as a constant-resistance load. This validates the definition of an anomalous resistivity $\eta \equiv E_{anom}/J$, which scales as $\eta \approx (B/n_e e)(1 - C_s/V_{d0})$ given the free energy source $J = n_e e V_{d0}$. To close the macroscopic transport model, we consider the generalized Einstein relation for non-equilibrium steady states, \cite{harada_equality_2005,loi_effective_2011} $D_{\text{eff}}\approx\mu k_BT_\text{eff}$, where $k_B$ is the Boltzmann constant. Since the turbulent fluctuations are spectrally limited by the thermal saturation scale ($C_s$), we adopt $T_\text{eff}\approx T_e$ and obtain $D_{\text{eff}} \approx T_e/n_e e^2 \eta$. For $\psi$, this yields the effective diffusion coefficient:
\begin{equation} \label{eq:bohm}
D_{\parallel} = \frac{T_e}{eB} \left( 1 - \frac{C_s}{V_{d0}} \right)^{-1} \xrightarrow{V_{d0} \gg C_s} \frac{T_e}{eB}.
\end{equation}
Thus, the friction generated by the electron slip stream is identified as the anomalous transport coefficient `\textit{Bohm'}, \cite{kaufmanExplanationBohmDiffusion1990,treumann_advanced_1997,burch_electron-scale_2016,braginskiiTransportProcessesPlasma1965,ottDiffusionStronglyCoupled2011} a phenomenological staple in plasma studies.
We can use Eq.~(\ref{eq:bohm}) to demarcate the theory's area of validity. We write $L_c \approx (2\sqrt{2}\pi/0.1C_s) (T_e/eB)$. With $T_e\sim0.1$~eV and $B\sim50,000$~nT, we obtain $L_c\sim350$~m. For scales $L\lesssim350$~m, the effective field theory is valid, and the stochastic sum total of polarization electric fields renormalize into Bohm resistivity in the electron slip stream, in the Hall currents.

To provide empirical evidence for this position, consider the observations in Figures~\ref{fig:swarmcomp} and \ref{fig:kinetic}d. A powerlaw in wavenumber with a steep exponent (spectral index) of $-8/3$ is associated with kinetic magnetohydrodynamic (MHD) turbulence, in particular kinetic Alfv\'{e}n waves (KAWs), \cite{boldyrev_spectrum_2012,chenKineticAlfvenWave2013,galtier_entanglement_2015,david_k_perp_2019} generated in the warm magnetospheric plasma. As these structures propagate into the cold, collisional ionosphere, they transition into inertial or dispersive Alfv\'{e}n waves,  acting as the structured driver ($V_{d0}$) for the ensemble polarization electric field (Eq.~\ref{eq:eanom}). Because the electron fluid is in a critical state of thermal saturation, the anomalous response field $E_{\text{anom}}$ is instantly generated to maintain marginal stability. This anomalous field forces the electron transport to strictly track the spatio-temporal structure of the magnetospheric driver ($V_{d0}$). Consequently, the distribution of turbulent parcels match the magnetospheric $k^{-8/3}$ scaling directly, which is what we are seeing in Figures~\ref{fig:swarmcomp}a,b) and \ref{fig:kinetic}a--d) (see also the Supplementary Materials). With the reservation that matching spectral indices does not imply a causality, we note that, within the established paradigm, there are no reasons to expect any signature of KAWs in the E-region plasma. More generally, there is likewise no reason to expect an overall steep (dissipative) spectrum for the E-region over four orders of magnitude in scale-size.

We next follow our effective field theory to its logical conclusion. In the macroscopic limit, the phase dynamics of the overdamped turbulent waves reduce to the Adler equation for synchronization phenomena \cite{adler_study_1946}. Consequently, we recover the observed linear response (Eq.~\ref{eq:scaling}),
\begin{equation} \label{eq:theory}
n \propto \frac{1}{2\lambda_0^2}\Delta E,
\end{equation}
where $\lambda_0$ is the effective damping at the dissipative fixed point, which retains the Bohm scaling, $\lambda_0 = D_\text{eff}k^2=(4\pi^2/9)(T_e/eB)$ at the radar wavelength. This response matches the empirical result (Figure~\ref{fig:kinetic}g), allowing us to identify the system's behavior with the current-voltage (``I-V'') characteristics of an overdamped Josephson junction. \cite{stewart_current-voltage_1968,ambegaokar_voltage_1969,wiesenfeld_synchronization_1996} In this analogy, the transition from a static `locked' state to a phase-slipping `running' state corresponds to the onset of the electron slip stream ($V_{d0} > C_s$). Crucially, the response in Figure~\ref{fig:kinetic}g exhibits the signature of a \textit{noise-broadened bifurcation} \cite{ambegaokar_voltage_1969,danner_injection_2021}, confirming that the onset of the `running' state, $V_d>C_s$, is a critical point created by the renormalized \textit{noise} of the ensemble's saturation electric fields.


The Adler-Ohmic bifurcation denotes a characteristic behaviour in physical systems where oscillating constituents create a voltage drop against an effective tilted washboard potential, one that facilitates a `running', or phase slipping, state, following a threshold of disorder. \cite{ambegaokar_voltage_1969} That we are able to observe this macroscopic description in the particular magnetosphere-ionosphere coupling created by Farley-Buneman waves, implies that the constituents of \textit{that} system likewise lock or slip against an effective tilted washboard potential. Subsequently, the magnetosphere-ionosphere coupling near intense diffuse aurorae is kept in a critical state sustained by efficient, turbulent energy dissipation, and its macroscopic description is ultimately governed by `Model A' universal dynamics, \cite{hohenberg_theory_1977} a class of purely dissipative systems.


While the foregoing must be substantiated and observed in closer detail, the immediate implications can be elucidated. The notion that the auroral ionosphere near diffuse aurorae self-organizes yields immediate explanatory power for the magnetosphere-ionosphere coupling. The driven system evolves toward a state of maximum energy dissipation, governed by Alfv\'{e}nic impedance matching and wave reflection. Precipitation that fails to reach the altitude with the lowest instability threshold \cite{st-mauriceNarrowWidthFarleyBuneman2023} may cause impedance mismatch for the incoming Alfv\'{e}n waves, causing Alfv\'{e}nic power to be reflected rather than dissipated. Consequently, the coupled system selectively sustains, and favours, precipitating electrons whose emission profile peak at altitudes around 105~km  where the FB instability threshold is lowest, offering a thermodynamic explanation for the preferred altitude of pulsating aurorae. \cite{hosokawaIonosphericVariationPulsating2015,partamiesOccurrenceAverageBehavior2017,dahlgren_variations_2017,kawamuraEstimationEmissionAltitude2020,tesema_observations_2020,nishimuraDiffusePulsatingAurora2020} What is more, non-linear wave heating suppresses dissociative recombination rates ($\propto T_e^{-1/2}\;$  \cite{schunkIonospheresTerrestrialPlanets1980}), and the subsequent net increase in conductance may provide concrete explanatory value to the problem of the ``saturating polar cap potential'' (see, e.g., Ref.~[\cite{siscoe_transpolar_2004}]).

Furthermore, when saturation feedback effectively governs the system, the turbulence evolves without hysteresis, providing an explanation for observations of directly driven auroral plasma turbulence with zero or near-zero lag, \cite{shen_magnetospheric_2024,ivarsen_characteristic_2025,brindley_intense_2025} as well as justification for the omission of hysteresis in global turbulence models (e.g., Ref.~[\cite{wiltbergerEffectsElectrojetTurbulence2017}]). 

The mechanism that achieves the foregoing offers the most penetrating insight, represented by the derivation of complete, closed-form macroscopic transport relations (culminating in Eq.~\ref{eq:bohm}). They demonstrate that, in two-stream plasmas, whenever the driving field attempts to push the drift velocity beyond the sound barrier, the system automatically generates Bohm resistivity to clamp the flow to the ion acoustic speed. This is accomplished through saturation electric fields, in a stochastic ensemble. \textit{The FB turbulence generates exactly enough noise, or effective temperature ($T_\text{eff} \approx T_e$) to assemble a thermal diffusion process at the acoustic scale.} This provides a parameter-free constitutive relation for modeling anomalous currents and heating in global ionospheric simulations.

\vspace{24pt}
\section{Methodology}

\subsection{Observational methodology}

This section presents the methodology in detail. First, this comes in the form of a summary of two recent papers of a highly technical nature, pertaining to coherent radar data from Ionospheric Continuous-wave E region Bistatic Experimental Auroral Radar (\textsc{icebear}) \cite{ivarsenDistributionSmallScaleIrregularities2023} and GPS signal analysis from the Canadian High-Arctic Ionospheric Network (\textsc{chain}). \cite{jayachandranCanadianHighArctic2009,song_investigation_2025} We then comment on current interpretations of spectral density measurements in the auroral ionosphere, before we present the application of renormalization group theory to the Farley-Buneman (FB) dispersion relation, constituting an effective field theory for FB turbulence.



\textit{ICEBEAR} -- We analyze coherent scatter radar data from \textsc{icebear}, an experimental radar capable of imaging the distribution of small-scale (3~m) plasma turbulence in 3-dimensions (3D), \cite{huyghebaertICEBEARAlldigitalBistatic2019,lozinskyICEBEAR3DLowElevation2022} yielding the \textsc{icebear}~3D dataset. The echoes are seen towards the northern horizon in Saskatchewan, Canada. The radar can record thousands of echo locations per second, yielding exceedingly large point-cloud datasets, \cite{ivarsenAlgorithmSeparateIonospheric2023,ivarsen_point-cloud_2024} and we are here segmenting the data in 6~second bins. The radar echoes observed inside each bin are clustered with the algorithm described in Ref.~[\cite{ivarsen_point-cloud_2024}], yielding clusters such as the one presented in Figure~\ref{fig:ice}a).

Ref.~[\cite{ivarsenDistributionSmallScaleIrregularities2023}] developed a method to correlate the spatial positions of such radar echo point-clouds using the two-point correlation function, $\xi(r)$,
\begin{equation} \label{eq:twopoint}
    n^2[1 + \xi(r)] = \langle \rho(x)\rho(x+r)\rangle,
\end{equation}
where $n$ is the average number density of echoes in a given volume, and $\rho(x)$ is the number density of echoes at location $x$, and $r$ is a distance away from $x$, and in Figure~\ref{fig:ice}b) we show $\xi(r)-1$ calculated for the cluster shown in Figure~\ref{fig:ice}a). The method yields spatial power spectra $P(k)$ through a Hankel transform, \cite{ivarsenDistinguishingScreeningMechanisms2016,baddourChapterTwoDimensionalFourier2011}
\begin{equation} \label{eq:solvethis}
   P(k) = \int\displaylimits_{0}^\infty\! \xi(r) J_0(kr) \; r dr,
\end{equation}
where $J_0(kr)$ are zeroth order Bessel functions of the first kind. Such power spectra yield spectral information roughly between the scale-sizes $750$~m and  $\sim10^5$~m, much larger than the \textsc{icebear} radar wavelength ($3$~m). Later, the spectra were demonstrated to match the small-scale structuring of field-aligned electrical currents associated with pulsating aurorae \cite{ivarsen_turbulence_2024-1} as well as F-region plasma structuring. \cite{ivarsenMeasuringSmallscalePlasma2023}

\begin{figure*}
    \centering
    \includegraphics[width=.95\textwidth]{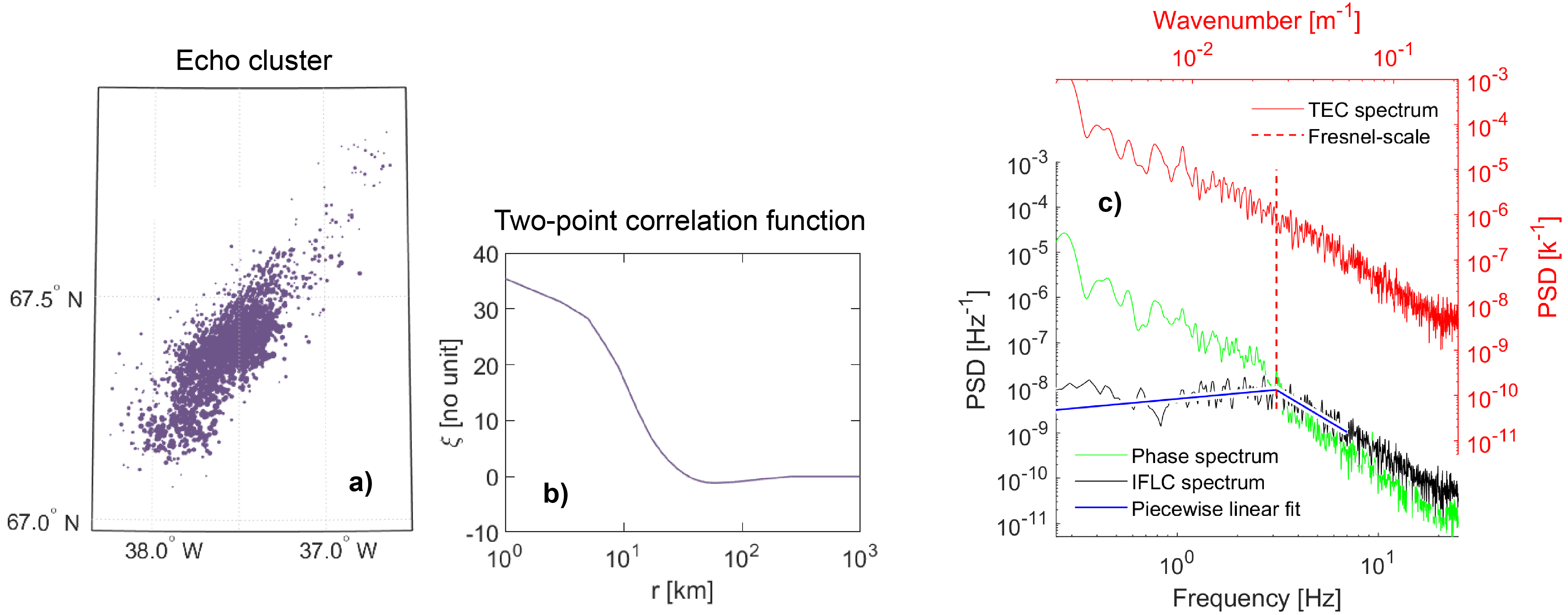}
    \caption{\textbf{Panel a)}  shows a sample 6-second radar point-cloud cluster,  while \textbf{Panel b)} shows the two-point correlation function based on that point-cloud (Eq.~\ref{eq:twopoint}) \textbf{Panel c)} shows example phase (green), IFLC (black), and TEC (red) spectra, with a piecewise linear fit shown with a blue line, and with the Fresnel scale (wavenumber) shown with a red, dashed line.}
    \label{fig:ice}
\end{figure*}

\textit{CHAIN} -- Then, we perform a time-series analysis of GPS amplitude and phase fluctuations measured with the  Ionospheric Scintillation Monitor Receiver (ISMR) located at the Rabbit Lake research station (58.23$^\circ$N, 256.32$^\circ$E), being part of the Canadian High Arctic Ionospheric Network  (\textsc{chain} \cite{jayachandranCanadianHighArctic2009}). The ISMR in Rabbit Lake is a Septentrio PolaRxS, \cite{bougard_cigala_2011} capable of recording the raw phase and post-correlation in-phase (I) and quadrature (Q) samples of GNSS signals at a 100~Hz sampling rate. For the purposes of our work, we concentrate solely on GPS.

The signals emitted by GPS satellites and subsequently recorded by the ISMR are disrupted by plasma irregularities linked with instabilities and turbulent phenomena in the ionospheric plasma, \cite{yehRadioWaveScintillations1982,kintnerp.m.GPSIonosphericScintillations2007} which introduce stochastic fluctuations in the recorded signal amplitude. \cite{mccaffrey_determination_2019,spogli_adaptive_2021,ghobadi_disentangling_2020,songSpectralCharacteristicsPhase2023,mezianeTurbulenceSignaturesHighLatitude2023} We  derive a spatial $k$-spectrum from the temporal fluctuations in the recorded signal in the Rabbit Lake ISMR (directly beneath the radar field-of-view) using phase screen theory. \cite{yehRadioWaveScintillations1982} In this framework, the Fresnel frequency $f_F$ is derived from the second zero-crossing of the normalized cross-spectrum of the L1-L2 or L1-L5 carrier frequencies, \cite{song_investigation_2025} or, equivalently, the prominent breakpoint or ``knee'' in the ionosphere-free linear combination (IFLC) spectrum, \cite{song_investigation_2025}, shown with a blue line in Figure~\ref{fig:ice}c). We then estimate the average drift velocity of the irregular structures in the F-region,  \cite{forte_problems_2002,spogli_adaptive_2021} $v_d$, using $v_d=f_F\lambda_F$, where $\lambda_F$ is the corrected Fresnel scale, \cite{ghidoni_ionospheric_2025}
\begin{equation}
\lambda_F = \sqrt{\frac{2\lambda_{\text{GPS}}h}{\sin \theta}},
\end{equation}
where $\lambda_{\text{GPS}} \approx~20$~cm is the wavelength of the GPS signal, $h$ is the altitude of the irregularity layer -- here assumed to be around $h=105$~km -- and $\theta$ is the GPS satellite elevation angle.  The expression for $\lambda_F$ takes into account the oblique incidence observational geometry, for which the distance between the antenna and the irregularity layer becomes $h\sin\theta$ and the cross-section of the irregularity, under the isotropic approximation, appears elliptical rather than circular. The latter introduces an additional azimuthal dependence due to the apparent major axis length relative to that of an ideal circular irregularity.  \cite{teunissen_springer_2017} As mentioned, we keep the estimation of the irregularity velocity under the approximation of a single, thin, isotropic irregularity layer. The velocity measured with this method yields an estimate of the relative velocity between the satellite pierce-point and the ionospheric irregularities, often referred as ``scan velocity.'' \cite{yehRadioWaveScintillations1982}

We construct a spectrum based on the IFLC spectrum (for scale-sizes smaller than the Fresnel scale), and a spectrum of Total Electron Content (TEC) fluctuations for scale-sizes larger than that threshold. The result is a $k$-spectrum that yields spectral information on scales roughly between 5~km and $\sim20$~m (the red spectrum in Figure~\ref{fig:ice}c), which is readily compared to the spectrum of ``echo clustering'' of co-located irregular structures seen by radar.

The GPS-derived scan velocity, on the other hand, is readily compared to the radar Doppler speeds (Figure~\ref{fig:kinetic}). However, the physical implications of the radar Doppler speeds hinge on the fact that the E-region plasma is highly collisional and is therefore not following the general $\boldsymbol{E}\times\boldsymbol{B}$-drift, and the true motion of this retarded flow is constrained to, roughly, the local ion acoustic speed  (see, e.g., Figure 2 in Ref.~[\cite{fosterSimultaneousObservationsEregion2000}]). We suffice here to write that the observed GPS scan velocities were highly consistent with observations of the retarded flow. This result is consistent across the conjunctions and justifies the assumption that the GPS pierce-points were concurrent with the observed E-region turbulence in altitude.

\begin{figure}
    \centering
    \includegraphics[width=0.495\textwidth]{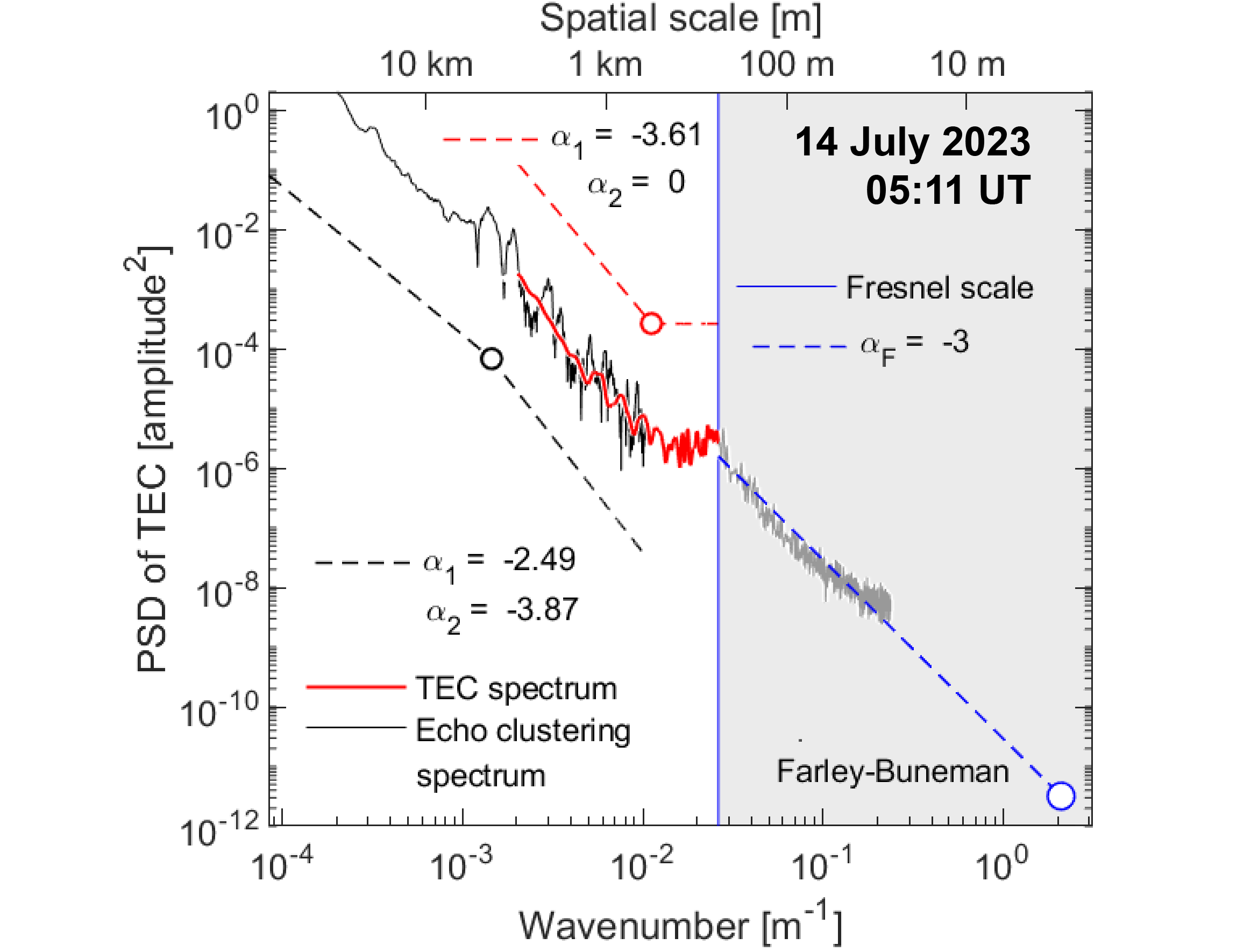}
    \caption{A conjunction between \textsc{icebear} and \textsc{chain}, yielding the \textit{composite spectrum}. Grey shading corresponds to the spectral component at scales smaller than the Fresnel scale, red line corresponds to the GNSS spectrum above the Fresnel scale, and black line corresponds to the \textsc{icebear} spectrum. Note the strong correspondence between the GNSS- and \textsc{icebear}-measured spectral slopes in the overlapping part of the spectrum (red $\alpha_{1,2}$ and black $\alpha_{1,2}$).}
    \label{fig:compspectrum}
\end{figure}

\subsubsection{Spectral density interpretation}

The result from combining the two methods of spectral density measurements described in the foregoing section is presented in Figure~\ref{fig:compspectrum}, in the form of the \textit{composite spectrum}, the primary quantity that is analyzed in the present study. The clustering spectrum (black) is seamlessly consistent with the TEC spectrum (red and gray). Prominent breakpoints are seen at spatial scales around 7~km and 300~m, as expected from the literature, and the inferred spectral index (see Eq.~\ref{eq:P}) is similar for the largest ($>50$~km) and smallest ($<100$~m) scale-sizes. What follows is a reflection on the role of such spectral density measurements in ionospheric plasma.

In studies of space plasma turbulence, one expects spectral densities to adhere to a simple power law,   \cite{phelpsPlasmaDensityIrregularities1976,tsunodaHighlatitudeRegionIrregularities1988,spicherDirectEvidenceDoubleslope2014}
\begin{equation} \label{eq:P}
    P(k) \propto k^{-\alpha},
\end{equation}
where $P(k)$ is the power spectral density, $k$ denotes wavenumber ($k=2\pi/L$, $L$ being spatial scale), and $\alpha$ is a positive constant describing the decay in spectral power with decreasing spatial scale (increasing $k$). \cite{sudanGenerationSmallscaleIrregularities1973,mounirSmallscaleTurbulentStructure1991,ivarsenSteepeningPlasmaDensity2021} $\alpha$ is referred to as spectral index and is a central quantity of measurement in the present study. In the auroral region of Earth's ionosphere, the steady decay in power indicated by Eq.~(\ref{eq:P}) is in fact often broken into segments, with typical `break-points' occurring on spatial scales between 1~km -- 5~km \cite{villainARCAD3SAFARICoordinatedStudy1986,ivarsen_what_2024,ivarsenMeasuringSmallscalePlasma2023} and between 30~m -- 300~m. \cite{basuPlasmaStructuringGradient1990,mounirSmallscaleTurbulentStructure1991,spicherDirectEvidenceDoubleslope2014,hamzaTwocomponentPhaseScintillation2023}

Break-points are often thought of as transition markers between inertial and fully collisional regimes. \cite{keskinenNonlinearEvolutionHighlatitude1990,kivancSpatialDistributionIonospheric1998} In the reigning view of ionospheric plasma turbulence, the various spectral indices, or slopes, and their relative magnitudes, are indicative of the instability processes that contribute to the structuring of the plasma. \cite{sudanGenerationSmallscaleIrregularities1973,mounirSmallscaleTurbulentStructure1991}

Another view posits that measured spectral density is produced by the dimensionality constraints that are placed on the system, in which the emergence of ordered dissipative structures are as much a \textit{consequence} and a \textit{source} of chaos, \cite{marov_self-organization_2013} and where spectral shape is dictated by \textit{self-organized criticality}, \cite{marek_can_2022} from which all the structure of the universe can presumptively be deduced.

\subsection{Renormalization group theory} \label{sec:rg}

To determine the macroscopic universality class of the system, we subjected the turbulent waves to a dynamic renormalization group analysis (see, e.g., Refs.~[\cite{wilson_renormalization_1971,hohenberg_theory_1977}]).  In the following, We construct an effective action using the Martin-Siggia-Rose formalism and apply to it a dynamic scaling analysis with spatial rescaling $x'=b^{-1}x$ and temporal rescaling $t'=b^{-z}t$. We then derive the macroscopic ``Adler-Ohmic'' linear power law observed in the auroral E-region ionosphere, from the microscopic physics of FB turbulence. The derivation is performed  in the neutral atmosphere's frame of reference, but an equivalent derivation in the ions' frame of reference yields identical results.

\subsubsection{The dispersion relation}

We write out the dispersion relation in the atmosphere frame, defining the wave frequency $\omega$ relative to the neutral gas,
\begin{equation} \label{eq:fulldisp}
    \omega^2 - k_\perp^2 C_s^2 + i(\omega - \mathbf{k} \cdot \mathbf{V}_{d0}) \gamma_\text{eff} = 0,
\end{equation}
where $\mathbf{V}_{d0}$ is the mean $\mathbf{E} \times \mathbf{B}$ drift velocity. We note that the real part of the frequency corresponds to the convective derivative in the time domain, while the imaginary damping term acts on the Doppler-shifted frequency.

Next, we shall apply the Martin-Siggia-Rose formalism of describing statistical dynamics for classical systems \cite{phythian_further_1976,andersen_functional_2000}, turning the stochastic differential equations of motion for our system into an effective, macroscopic field theory \cite{bonicelli_algebraic_2025}. We start by constructing the unperturbed action in the path integral formalism, based on Eq.~(\ref{eq:fulldisp}). In the neutral gas frame, this explicitly retains the mean advection of the plasma structures:
\begin{multline} \label{eq:unperturbed}
    S_0 = \int dt \, d^d x \, \tilde{\psi} \left[\partial_t^2 + 2(\mathbf{V}_{d0} \cdot \nabla)\partial_t + (\mathbf{V}_{d0} \cdot \nabla)^2 - \right. \\ \left. - C_s^2 \nabla^2 + \gamma_\text{eff} (\partial_t + \mathbf{V}_{d0} \cdot \nabla) \right] \psi,
\end{multline}
where $\psi=\delta N_e/N_{e,0}$ is the electron density variation divided by mean density, and $\tilde{\psi}$ is the auxiliary response field acting as a Lagrange multiplier to enforce the deterministic trajectory \cite{bonicelli_algebraic_2025}. A two-step argument treats the non-linearities in the turbulent plasma:

\textbf{\textit{(1)}} The linear propagator $S_0$ accounts for the mean flow but neglects the effects of turbulent mode-coupling. However, the physical mechanism at the level of individual structures involves the development of secondary polarization fields. A justification for treating these non-linearities stochastically can be derived from the work of Refs.~[\cite{st.-mauriceNewNonlinearApproach2001,drexlerNewInsightsNonlocal2002}]. Based on Fourier analysis, Ref.~[\cite{st.-mauriceNewNonlinearApproach2001}] showed that for an initial finite-size structure, polarization electric fields are created that oppose the driving field. These internal fields saturate the instability locally, explaining why VHF Doppler speeds cluster around $C_s$ (Figure~\ref{fig:kinetic}b).

\textbf{\textit{(2)}} As a result of point~\textit{(1)}, we construct a stochastic source term driven by these self-generated polarization fields. Rather than a slow modulation, we recognize that the total drift velocity $\mathbf{V}(x,t)$ at any point is the sum of the coherent mean drift $\mathbf{V}_{d0}$ and a fluctuating component $\delta \mathbf{v}(t)$ arising from the turbulent ensemble:
\begin{equation}
    \mathbf{V}(x,t) = \mathbf{V}_{d0} + \delta \mathbf{v}(x,t).
\end{equation}
Because these fluctuations are driven by the instability itself, they occur on fast timescales, and we write the time-dependent drift for an individual structure $i$ as
\begin{equation}
    \mathbf{V}_i(t) = \mathbf{V}_0 \left[ 1 - \varepsilon \sin(\Omega t + \alpha_i) \right],
\end{equation}
where $\Omega$ represents the waves' life-cycle, $\alpha_i$ represents the random phase of the $i^\text{th}$ structure, and $\varepsilon$ is the amplitude of the non-linear feedback. Furthermore, since the macroscopic turbulent electrojet is an ensemble of such structures spread over $10$s of km, the phase information of the fluctuations is effectively random across the volume. We therefore define the stochastic velocity field $\delta \mathbf{v}(x,t)$ via the random phase approximation (RPA) \cite{krommes_fundamental_2002},
\begin{equation}
    \delta \mathbf{v}(x,t) = - \varepsilon \sum_i \mathbf{v}_i \sin(\Omega t + \alpha_i(x)).
\end{equation}
By virtue of the central limit theorem, the sum of these uncorrelated, fast-oscillating terms converges to a Gaussian random variable. Thus, we formally replace the deterministic non-linear trajectory with a stochastic field $\delta \mathbf{v}$ governed by a Gaussian probability distribution $P[\delta \mathbf{v}]$,
\begin{equation}
    P[\delta \mathbf{v}] \propto \exp\left( -\int dt \, d^d x \frac{|\delta \mathbf{v}|^2}{2\sigma_v} \right).
\end{equation}
Here, the variance $\sigma_v$ is constrained by the power of the turbulent electric fields. Following the MSR formalism, this stochastic noise enters the equation of motion as a multiplicative source term acting on the gradient of the density field,
\begin{equation}
    \hat{\mathcal{L}}_0 \psi = -\gamma_\text{eff} (\delta \mathbf{v} \cdot \nabla) \psi,
\end{equation}
which introduces a linear coupling between the response field $\tilde{\psi}$ and the noise ensemble,
\begin{equation}
    S_{\text{coupling}} = \int dt \, d^d x \, \tilde{\psi} \left[ \gamma_\text{eff} (\delta \mathbf{v} \cdot \nabla) \psi \right].
\end{equation}
To derive the effective macroscopic action, we integrate over the ensemble of phase histories. We compute the expectation value of the coupling term over the Gaussian distribution,
\begin{equation}
    I = \left\langle \exp\left( i \int dt \, d^d x \, (\gamma_\text{eff} \tilde{\psi} \nabla \psi) \cdot \delta \mathbf{v} \right) \right\rangle_{\delta \mathbf{v}}.
\end{equation}
Using the Gaussian integral identities, we obtain,
\begin{equation}
    I = \exp\left( -\frac{\sigma_v}{2} \int dt \, d^d x \left( \gamma_\text{eff} \tilde{\psi} \nabla \psi \right)^2 \right).
\end{equation}
The MSR path integral weight is defined as $\mathcal{Z} = \int \mathcal{D}\psi \mathcal{D}\tilde{\psi} \, e^{i S}$. To translate the real-valued Gaussian probability weight of the noise fluctuations into the complex-weighted path integral formalism, we identify the integration result with $iS_\text{int}$, the imaginary component of the action,
\begin{equation}
    \label{eq:stochvertex}
i S_{\text{int}} = -\frac{\sigma_v}{2} \gamma_\text{eff}^2 \int dt \, d^d x \, (\tilde{\psi} \nabla \psi)^2.
\end{equation}

We can now combine the deterministic propagator (derived from the dispersion relation, Eq.~\ref{eq:fulldisp}), and the stochastic noise term, (Eq.~\ref{eq:stochvertex}), obtaining the total action, \begin{multline} \label{eq:action} S = \int dt , d^d x \left\{ \tilde{\psi} \left[ \partial_t^2 + 2(\mathbf{V}_{d0} \cdot \nabla)\partial_t + (\mathbf{V}_{d0} \cdot \nabla)^2 - \right. \right. \\ - \left. \left. C_s^2 \nabla^2 + \gamma_\text{eff} (\partial_t + \mathbf{V}_{d0} \cdot \nabla) \right] \psi - i \frac{\sigma_v}{2} \gamma_\text{eff}^2 (\tilde{\psi} \nabla \psi)^2 \right\} \end{multline}

\subsubsection{Renormalization group flow}

We determine the macroscopic phase of the system by analyzing the flow of the coupling constants under a rescaling of spacetime,
\begin{align*}
x &\to b x, \\
t &\to b^z t, \\
\psi &\to b^{\chi} \psi, \\
\tilde{\psi} &\to b^{\tilde{\chi}} \tilde{\psi}.
\end{align*}
The integration measure therefore transforms as $\int dt \, d^d x \to b^{z+d} \int dt \, d^d x$. We next choose the stiffness operator (diffusion) to be scale-invariant. In other words, with the scaling relations applied to $C_s^2 \tilde{\psi} \nabla^2 \psi$ (Eq.~\ref{eq:action}), we obtain the constraint,
\begin{equation} \label{eq:chiscale}
    \chi + \tilde{\chi} = 2 - z - d.
\end{equation}
Applying this constraint to the other operators in the action to find their scaling behavior relative to the stiffness, we find that,
\begin{align}
    \int dt \, d^d x \, \tilde{\psi} \partial_t^2 \psi \; \sim \; b^{(2-z-d) + z + d - 2z} &= b^{2-2z}, \label{eq:scale1} \\
    \int dt \, d^d x \, \gamma_\text{eff} \tilde{\psi} \partial_t \psi \; \sim\; b^{(2-z-d) + z + d - z} &= b^{2-z}, \\
    \int dt \, d^d x \, \gamma_\text{eff} \tilde{\psi} (\mathbf{V}_{d0} \cdot \nabla) \psi \; \sim\; b^{(2-z-d) + z + d - 1} &= b^{1}, \label{eq:scale_advection} \\
    \int dt \, d^d x \, \tilde{\psi} (\mathbf{V}_{d0} \cdot \nabla)^2 \psi \; \sim\; b^{z+d+\chi+\tilde{\chi}-2} &= b^0,  \\
    \int dt \, d^d x \, \frac{\sigma_v}{2} \gamma_\text{eff}^2 (\tilde{\psi} \nabla \psi)^2 \; \sim\; b^{z+d - 2 + 2(\chi+\tilde{\chi})}  &= b^{2 - z - d},
\end{align}
where we used Eq.~(\ref{eq:chiscale}). We observe that at the inertial fixed point ($z=1$), the inertial term (Eq.~\ref{eq:scale1}) is relevant. However, with increasing $z$ the system flows to the dissipative fixed point at $z=2$, where the inertial term becomes irrelevant ($b^{-2}$) and the damping term becomes marginal ($b^0$). Crucially, the linear advection term (Eq.~\ref{eq:scale_advection}) scales as $b^1$, identifying it as a relevant operator: macroscopic transport is dominated by the mean drift.

In the macroscopic limit, the inertial terms ($\partial_t^2$) and the stochastic driver fluctuations ($\sigma_v$) are renormalized. The stochastic variance $\sigma_v$ generates the anomalous diffusion tensor, while the mean drift remains as a relevant transport term. This leaves a deterministic effective theory describing an advecting, diffusing fluid,
\begin{equation} \label{eq:action_final}
    S = \int dt \, d^d x  \tilde{\psi} \left[ \gamma_\text{eff} (\partial_t + \mathbf{V}_{d0} \cdot \nabla) - (\mathbf{V}_{d0} \cdot \nabla)^2 - C_s^2 \nabla^2 \right] \psi. 
\end{equation}

At this point, we reinstate the internal feedback reaction that the unstable structures excite \cite{st.-mauriceNewNonlinearApproach2001}, namely that the turbulent electric field is periodically brought down to the threshold $\mathbf{V}_{d0}\to\mathbf{V}_d=\mathbf{V}_{d0}\left(1 - \tilde{\alpha} \langle |\psi|\rangle^2 \right)$, where we recognize that the variance of the microscopic velocity fluctuations ($\sigma_v$) manifests macroscopically as an anomalous drag force. The feedback follows: as the turbulent amplitudes increase, the driving electric field is brought down towards $C_s$, which in turn kills the instability and lowers the turbulent amplitude ($\langle |\psi|\rangle^2 \to0$). As we shall demonstrate, this description ensures that the system finds the fixed point where the effective drift is just above the ion acoustic speed (Figure~\ref{fig:kinetic}b). We obtain the deterministic action,
\begin{equation} \label{eq:action}
    S = \int dt \, d^d x  \tilde{\psi} \left[ \gamma_\text{eff} (\partial_t + \mathbf{V}_d \cdot \nabla) - (\mathbf{V}_d \cdot \nabla)^2 - C_s^2 \nabla^2 \right] \psi. 
\end{equation}
Here, we can combine the two stiffness terms by way of tensor notation,
\begin{equation}
    \left[ (\mathbf{V}_d \cdot \nabla)^2 - C_s^2 \nabla^2 \right] \psi = -\nabla \cdot \left( C_s^2 \mathbb{I} - \mathbf{V}_d \otimes \mathbf{V}_d \right) \cdot \nabla \psi,
\end{equation}
which allows us to write the first-order, overdamped advection-diffusion equation,
\begin{multline} \label{eq:transport1}
    S_{\text{eff}} = \int dt \, d^d x \, \tilde{\psi} \; \gamma_\text{eff}\left[  (\partial_t + \mathbf{V}_d \cdot \nabla) - \right. \\ \left. - \gamma_\text{eff}^{-1} \nabla \cdot \left( C_s^2 \mathbb{I} - \mathbf{V}_d \otimes \mathbf{V}_d \right) \cdot \nabla \right] \psi.
\end{multline}
We note that the spatial dynamics are governed by the effective diffusion tensor $\mathbf{D}_{\text{eff}} = \gamma_\text{eff}^{-1}(C_s^2 \mathbb{I} - \mathbf{V}_d \otimes \mathbf{V}_d)$, with $\mathbf{V}_d$ representing the saturated drift,
\begin{equation}
    \mathbf{D}_{\text{eff}} = \frac{1}{\gamma_\text{eff}}\begin{bmatrix} C_s^2 - V_d^2 & 0 & 0 \\ 0 & C_s^2 & 0 \\ 0 & 0 & C_s^2 \end{bmatrix}.
\end{equation}

\textbf{\textit{(1)}} In directions parallel to the mean advection $\mathbf{V}_{d0}$, if the driving drift $V_{d0} > C_s$, the effective diffusion coefficient becomes \textit{negative}, which we identify as the macroscopic signature of the instability. This is followed by saturation,
\begin{equation}
D_{\parallel}(\psi) \approx (C_s^2 - V_{d0}^2) + 2\tilde{\alpha} V_{d0}^2 |\psi|^2,
\end{equation}
where we have linearized the effective drift squared ($V_d^2 \approx V_{d0}^2 - 2\tilde{\alpha}V_{d0}^2|\psi|^2$) and neglected higher-order $\mathcal{O}(|\psi|^4)$ terms. The negative diffusion that commences when $V_{d0} > C_s$ leads to perturbation growth, which in turn activates the anomalous drag (saturation) through $|\psi|^2$, stabilizing the transport.

\textbf{\textit{(2)}} In directions perpendicular to the drift, the diffusion simplifies to $\mathbf{D}_{\text{eff}} \approx C_s^2$, the transverse terms representing the ordinary diffusion of the transverse modes. These act as an energy sink, aligning with Ref.~[\cite{hamza_turbulent_1993}], who explicitly identified the mode-coupled secondary waves as acting akin to ``driven, overdamped oscillators.''

\subsubsection{Advective-diffusive dynamics} \label{sec:tensorhome}

The foregoing subsection established that at macroscopic scales ($z=2$), the inertial terms ($\partial_t^2$) are irrelevant. The system flows universally to an advective-diffusive state, and so we write the effective equation of motion for the turbulent density fluctuation field $\psi(\mathbf{x},t)$ as,
\begin{multline}
\gamma_\text{eff} \left( \frac{\partial}{\partial t} + \mathbf{V}_{d0} \cdot \nabla \right) \psi - \nabla \cdot \left[ (C_s^2 - V_{d0}^2) +  \right. \\ \left. + 2\tilde{\alpha} V_{d0}^2 |\psi|^2 \right] \cdot \nabla \psi = S(\mathbf{x},t), 
\end{multline}
where $\gamma_\text{eff}$ is the renormalized effective damping. $S(\mathbf{x},t)$ is a source term,  
which we decompose into an amplitude $\mathcal{E}$ and a phase $\theta(t)$,
\begin{equation} \label{eq:source}
    S(\mathbf{x},t) = \mathcal{E} \, e^{i(\mathbf{k} \cdot \mathbf{x} - \theta(t))},
\end{equation}
where $\mathcal{E}$ is the strength of the driver and $\theta$ is phase. Here, we consider the VHF radar's (and the instability's) natural preference for the driving modes that resonate with the ion acoustic speed. To account for this observational effect, we approximate the driver phase speed with the acoustic speed characteristic of the saturated state,
\begin{equation}
    \dot{\theta} \approx k C_s.
\end{equation}

We next express the turbulent plasma density field $\psi$ in terms of a slowly varying amplitude $A(t)$ and phase $\phi(t)$:
\begin{equation} \label{eq:response}
    \psi(\mathbf{x},t) = A(t) \, e^{i(\mathbf{k} \cdot \mathbf{x} - \phi(t))}. 
\end{equation}
We substitute Eqs.~(\ref{eq:source}) and (\ref{eq:response}) into the effective equation of motion. The convective derivative on the left-hand-side expands as,
\begin{equation}
     \left( \frac{\partial}{\partial t} + \mathbf{V}_{d} \cdot \nabla \right) \psi =  \left[ (\dot{A} - i A \dot{\phi}) + i A (\mathbf{k} \cdot \mathbf{V}_{d}) \right] e^{i(\mathbf{k} \cdot \mathbf{x} - \phi)}.
\end{equation}
Combining this with the diffusion term,
\begin{equation} 
    -\nabla \cdot \mathbf{D}_{\text{eff}} \cdot \nabla \psi = \left( \lambda_0 + \tilde{\alpha} V_{d0}^2 |\psi|^2 \right) \psi, 
\end{equation}
where $\lambda_0$ is the positive, renormalized, anomalous diffusion coefficient, and where we neglect higher-order non-linear interaction terms. With Eq.~(\ref{eq:source}), the full macroscopic equation of motion reads,
\begin{equation} \label{eq:full0}
    \left[ \dot{A} - i A (\dot{\phi} - \mathbf{k} \cdot \mathbf{V}_{d}) \right] + \lambda_0 A = \mathcal{E} (\cos \Theta + i \sin \Theta).
\end{equation}
where we defined the phase mismatch $\Theta(t) \equiv \phi(t) - \theta(t)$. Finally, Eq.~(\ref{eq:full0}) can be expressed more sensibly by recognizing that,
\begin{equation}
\dot{\phi} - \mathbf{k} \cdot \mathbf{V}_{d} = \dot{\Theta} + \dot{\theta} - \mathbf{k} \cdot \mathbf{V}_{d} \approx  \dot{\Theta} - k(V_{d}-C_s),
\end{equation} 
which, when substituted this into Eq.~(\ref{eq:full0}), yields the final, macroscopic equation of motion,
\begin{equation} \label{eq:full}
    \left[ \dot{A} - i A \left( \dot{\Theta} - k(V_{d} - C_s) \right) \right] + \lambda_0 A = \mathcal{E} (\cos \Theta + i \sin \Theta).
\end{equation}

\textbf{\textit{(1)}} The imaginary part of Eq.~(\ref{eq:full}) reveals the phase dynamics relative to the mean flow. 
\begin{equation}
- A \left[ \dot{\Theta} - k(V_{d} - C_s) \right] = \mathcal{E} \sin \Theta,
\end{equation}
which we can rearrange to isolate $\dot{\Theta}$, yielding the Adler equation for phase-locking phenomena, \cite{adler_study_1946,ambegaokar_voltage_1969,wiesenfeld_new_1996}
\begin{equation} \label{eq:adlernew}
\dot{\Theta} = k(V_{d} - C_s) - \frac{\mathcal{E}}{ A} \sin \Theta.
\end{equation}

\textbf{\textit{(2)}} Taking the real part of Eq.~(\ref{eq:full}) yields,
\begin{equation}
 \dot{A} + \lambda_0 A = \mathcal{E} \cos \Theta,
\end{equation}
which we can solve for the steady state by applying singular perturbation theory \cite{arnold_geometric_1995}. The characteristic lifetime of FB turbulence is \textit{short}, on millisecond timescales ($<1$~s), while the driver $\mathcal{E}$ varies over seconds, enabling us to write $\epsilon = \tau/T_{\text{driver}} \ll 1$. In the limit $\epsilon \to 0$, the time derivative $\epsilon \dot{A}$ vanishes. The system collapses onto the slow manifold,
\begin{equation} \label{eq:slowmani}
A(t) \approx \frac{\mathcal{E}}{\lambda_0} \cos \Theta(t).
\end{equation}
To relate this to the radar observations, we approximate the echo intensity $n$ with the turbulence power, or the time-averaged square of the amplitude, $n = \langle A^2 \rangle_t$. Squaring Eq.~(\ref{eq:slowmani}) and time-averaging over the rapid phase slippage (where $\langle \cos^2 \Theta \rangle = 1/2$), we find that the turbulence intensity becomes proportional to the square of the driver strength,
\begin{equation} \label{eq:scaling}
n \propto \frac{1}{2\lambda_0^2} \mathcal{E}^2.
\end{equation}
Since the source term $\mathcal{E}^2$ is proportional to the input wave power $\Delta E$, this strictly recovers the Adler-Ohmic regime observed in Figure~\ref{fig:kinetic}g): the turbulence intensity scales linearly with the driving power, confirming that the complex microscopic physics linearize in the macroscopic view of the system.

\subsubsection{The cross-over scale} \label{sec:rg:scale}

From Eq.~(\ref{eq:adlernew}), we define the phase-locking criterion,
\begin{equation} \label{eq:ineqalmostlast}
|k(V_{d} - C_s)| \leq \frac{\mathcal{E}}{ A},
\end{equation}
yielding the stability bandwidth. We derive the crossover-scale $L_c$ separating the inertial drift regime from the macroscopic effective field theory.
We utilize the slow-manifold solution to the amplitude equation, $A=\mathcal{E}/(\sqrt{2}\lambda_0)$. Substituting the renormalized diffusive linewidth $\lambda_0=D_\text{eff}k^2$ into Eq.~(\ref{eq:ineqalmostlast}) yields the scale-dependent locking condition:
\begin{equation} \label{eq:ineqlast}
k(V_{d}-C_s) \leq \sqrt{2}D_\text{eff}k^2.
\end{equation}
We solve for the critical wavenumber $k_c$, the minimum $k$ required for the diffusive damping (right-hand-side) to overcome the linear drift drive (left-hand side):
\begin{equation}
k_c = \frac{V_{d} - C_s}{\sqrt{2}D_\text{eff}},
\end{equation}
corresponding to the cross-over scale for the RG flow,
\begin{equation}
L_c = \frac{2\sqrt{2}\pi D_\text{eff}}{V_{d}-C_s}.
\end{equation}
This length scale demarcates the validity of the effective field theory. For small scales ($L < L_c$), the renormalized diffusion dominates, and the system is described by the macroscopic transport. In other words, the system's macroscopic correlation length is characteristic of a continuous phase transition, \cite{hohenberg_theory_1977}  diverging at $V_{d}=C_s$.

\subsubsection{Caveats \& kinetic theory} \label{sec:rg:kinetic}

The mathematical substitution of the deterministic mode-coupling term ($\psi \nabla \psi$) with a stochastic noise term ($\delta \mathbf{v}$) constitutes random phase approximation, which formally lumps the turbulent cascade into a diffusion term. By discarding the phase correlations required for three-wave interactions, this approach explicitly neglects the dynamic flux of energy across scales and treats the plasma as an ensemble of turbulent waves rather than a structured fluid capable of forming shocklets. This departure from a fluid description is validated by the two-stream origin of the instability: saturation is dominated by local electrostatic feedback (where polarization fields clamp the local drift to $C_s$), a notion that is firmly supported by Figure~\ref{fig:kinetic}b, as well as the state of electrojet turbulence research (see, e.g., Refs.~[\cite{fosterSimultaneousObservationsEregion2000,oppenheim_kinetic_2013,koustovRelationshipVelocityEregion2005,chauUnusualRegionFieldaligned2016}]).

A phenomenological argument based on the foregoing recovers the central result (Eq.~\ref{eq:scaling}), by way of adiabatic elimination. On time scales longer than the initial growth time, the unstable structures will lose amplitude after zero growth is approached: Ref.~[\cite{drexlerNewInsightsNonlocal2002}] showed that the eigenfrequency is a weak function of space (altitude), which forces the introduction of steadily increasing parallel electric fields inside unstable structures.  Such fields short-circuit the internal, perpendicular fields by passing parallel currents. \cite{oppenheim_kinetic_2013} This, incidentally, gives the wave energy to electrons which in turn give it to the gas through inelastic collisions. \cite{st-mauriceRevisitingBehaviorERegion2021}

Assume, then, that inside  an unstable volume, the faster the individual structures grow, the more of them there are. This introduces an equivalence between time and space: shorter  growth times mean more structures. \textit{The number of structures in an unstable volume is therefore proportional to the linear growth rate.} Using  fluid theory approximation as a guide this means that the number of structures inside the unstable volume is given by, 
\begin{equation}
    n \propto  a  \frac{\Psi}{1+\Psi}  \frac {k^2}{\nu_i}(C_s^2- V_d^2),
\end{equation}
where $a$ is a constant, $\nu_i$ the ion-neutral collision frequency, and $V_d$ is the relative ion-electron drift, very close to $E/B$ at 100 km altitude. Therefore, when $V_d ^2$ is well above $C_s^2$, the number of VHF radar backscatter targets becomes essentially proportional to the electric field squared, $E^2$.  With a dominance of the 500 to 600 m/s threshold speed observed  at 100 km,  E/B has to be exceeding 1500 m/s (see Figure~1 in Ref.~[\cite{st-mauriceNarrowWidthFarleyBuneman2023}]), and the near proportionality to $E^2$ is justified.


\section*{Acknowledgements}
This work is supported in part by the European Space Agency’s Living Planet Grant No. 1000012348 and by the Research Council of Norway (RCN) Grant No. 324859. We acknowledge the support of the Canadian Space Agency (CSA) [20SUGOICEB], the Canada Foundation for Innovation (CFI) John R. Evans Leaders Fund [32117], the Natural Science and Engineering Research Council (NSERC), the Discovery grants program [RGPIN-2019-19135], the Digital Research Alliance of Canada [RRG-4802].
DRT is supported through UK Natural Environment Research Council DRIIVE [NE/W003368/1] and FINESSE [NE/W003147/1] grants. Science data of the ERG (Arase) satellite were obtained from the ERG Science Center operated by ISAS/JAXA and ISEE/Nagoya University (\url{https://ergsc.isee.nagoya-u.ac.jp/index.shtml.en}. This includes Lv.3 HEP (DOI \texttt{10.34515/DATA.ERG-01002}), Lv.3 MEP-e (DOI \texttt{10.34515/DATA.ERG-02003}) and Lv.2 LEP-e (DOI \texttt{10.34515/DATA.ERG-05000}), Lv.2 PWE/OFA (DOI \texttt{10.34515/DATA.ERG-08000}), and Lv.2 MGF (DOI \texttt{10.34515/DATA.ERG-06001}). Super\textsc{mag} data can be accessed at \texttt{https://supermag.jhuapl.edu/mag/}.\textsc{icebear} 3D echo data for 2020, 2021 is published with DOI \texttt{10.5281/zenodo.7509022}. \textsc{chain} ISMR data is available at \url{https://chain-new.chain-project.net/index.php/data-products/data-download}. Google's Gemini 3.0 Pro has been used to assist mathematical formalism. MFI is grateful to M. Oppenheim, J. Park, and R. Horne for stimulating discussions.
\\
\\ 
See the Supplementary Materials [URL] for a complete and technical description of all the conjunction studies performed. The Supplementary Materials includes Refs.~[\cite{tsyganenko_modeling_2005,derrico_slm-shape_2009,parkAlfvenWavesAuroral2017,ivarsenObservationalEvidenceRole2020,sydorenko_stabilizing_2017,ghadjariStandingAlfvenWaves2022}].


%

\end{document}